\documentclass[%
 reprint,
 amsmath,amssymb,
 aps,
]{revtex4-2}

\usepackage{graphicx}
\usepackage{dcolumn}
\usepackage{bm}
\usepackage{color,mathrsfs}
\usepackage{amsmath}
\usepackage{hyperref}

\newcommand{\mnras}{{Mon.~Not.~R.~Astron.~Soc.}}
\newcommand{\aap}{{A\&A}}
\newcommand{\apjs}{{Astrophys.~J.~Supp.}}
\newcommand{\aj}{{Astro.~Journal}}
\newcommand{\araa}{{Ann.~Rev.~Astron.~Astrophys. }}

\newcommand{\jcap}{{J. Cosmol. Astropart. Phys.}}
\newcommand{\physrep}{{Physics.Report.}}
\newcommand{\apjl}{{Astrophysics. Journal. Letters.}}

\def\lsim{\;\rlap{\lower 2.5pt
    \hbox{$\sim$}}\raise 1.5pt\hbox{$<$}\;}
\def\gsim{\;\rlap{\lower 2.5pt
    \hbox{$\sim$}}\raise 1.5pt\hbox{$>$}\;}

\begin{document}

\preprint{APS/123-QED}

\title{The effects of peculiar velocities on the morphological properties of large-scale structure}

\author{Aoxiang Jiang}
\email{jax9709@mail.ustc.edu.cn}
\affiliation{CAS Key Laboratory for Research in Galaxies and Cosmology, Department of Astronomy, University of Science and Technology of China, Hefei, Anhui, 230026, P.R.China}
\affiliation{School of Astronomy and Space Sciences, University of Science and Technology of China, Hefei, Anhui, 230026, P.R.China}

\author{Wei Liu}
\affiliation{CAS Key Laboratory for Research in Galaxies and Cosmology, Department of Astronomy, University of Science and Technology of China, Hefei, Anhui, 230026, P.R.China}
\affiliation{School of Astronomy and Space Sciences, University of Science and Technology of China, Hefei, Anhui, 230026, P.R.China}
\author{Wenjuan Fang}
\email{wjfang@ustc.edu.cn}
\affiliation{CAS Key Laboratory for Research in Galaxies and Cosmology, Department of Astronomy, University of Science and Technology of China, Hefei, Anhui, 230026, P.R.China}
\affiliation{School of Astronomy and Space Sciences, University of Science and Technology of China, Hefei, Anhui, 230026, P.R.China}
\author{Wen Zhao}
\affiliation{CAS Key Laboratory for Research in Galaxies and Cosmology, Department of Astronomy, University of Science and Technology of China, Hefei, Anhui, 230026, P.R.China}
\affiliation{School of Astronomy and Space Sciences, University of Science and Technology of China, Hefei, Anhui, 230026, P.R.China}

\date{\today}

\begin{abstract}

It is known that the large-scale structure (LSS) mapped by a galaxy redshift survey is subject to distortions by galaxies' peculiar velocities. Besides the signatures generated in common N-point statistics, such as the anisotropy in the galaxy 2-point correlation function, the peculiar velocities also induce distinct features in LSS's morphological properties, which are fully described by four Minkowski functionals (MFs), i.e., the volume, surface area, integrated mean curvature and Euler characteristic (or genus). In this work, by using large suite of N-body simulations, we present and analyze these important features in the MFs of LSS on both (quasi-)linear and non-linear scales, {with a focus on the latter}. 
We also find the MFs can give competitive constraints on cosmological parameters compared to the power spectrum, probablly due to the non-linear information contained. {For galaxy number density similar to the DESI BGS galaxies, the constraint on $\sigma_8$ from the MFs with one smoothing scale can be better by $\sim 50\%$ than from the power spectrum.} These findings are important for the cosmological applications of MFs of LSS, and probablly open up a new avenue for studying the peculiar velocity field itself. 

\end{abstract}

\maketitle

\section{\label{sec:intro} Introduction}

The Universe's large-scale structure contains a wealth of information about its origin and development, including the initial conditions \citep[see e.g.,][]{Kaiser84, Dalal+08}, the forms and amounts of the constituent energy components \citep[see e.g.,][]{Carroll+92, Hu+98}, the laws of gravity \citep[see e.g.,][]{Clifton+12,  Baker+19} etc., and has been extensively used to probe these questions that are clearly of fundamental importance to cosmology. Galaxy redshift surveys, which map the 3-dimentional distribution of galaxies, provide a most direct way to measure the Universe's large-scale structure, therefore are actively pursued by the community, such as the SDSS \citep{York+00}, WiggleZ \citep{2010MNRAS.401.1429D}, PFS \citep{Takada_2014}, DESI \citep{2016arXiv161100036D}, CSST \citep{CSST, 2019ApJ...883..203G}.

However, the large-scale structure mapped by galaxy redshift surveys is obscured by galaxies' peculiar velocities \citep[see][for a review]{1998ASSL..231..185H}, in addition to other systematic effects such as galaxy bias, the Alcock-Paczynski effect \citep{1979Natur.281..358A}. The line-of-sight (LOS) component of the velocity other than the Hubble flow introduces a correction to a galaxy's distance from us which is intepreted from its redshift. Thus, on large scales, an overdense region tends to be squashed along the LOS, known as the Kaiser's effect \citep{1987MNRAS.227....1K}, while on small scales, it tends to be enlongated along the LOS, known as the fingers-of-God (FOG) effect \citep{10.1093/mnras/156.1.1P}. On the other hand, this ``redshift distortion'' to the measured large-scale structure provides a unique way to probe the peculiar velocity field, which is applicable even at high redshifts as opposed to direct measurements of the velocity field \citep[e.g.,][]{Gorski+89, Huterer+17}. For example, the galaxy correlation function, which is no longer isotropic in redshift space, contains important 2-pt statistics for the velocity field from which we have constructed the most widely used method of measuring the growth rate of structure \citep[e.g.][]{PerWhi09, Alam+17}.

Peculiar velocities principally affect all statistical properties of galaxy distribution in the 3D redshift space. Besides the well-studied 2-pt statistics which is sufficient to describe the field on linear scales, N($>2$)-pt statistics are also essential to describe the field on non-linear scales where the field is no longer Gaussian. Generally speaking, analysis with these $N$-pt statistics with $N>2$ are complicated or even infeasible at the moment \citep[see e.g.,][]{GilMarin+15, GilMarin+17, 2018JCAP...12..035D}, so alternatives such as the morphorlogical descriptors of Minkowski functionals (MFs) \citep{Minkowski1903, Mecke+94, 1997ApJ...482L...1S, SchGor98} have been proposed.

According to Hadwiger's theorem \citep{Hadwiger57, rado1959h}, for a spatial pattern in 3D, its morphorlogical properties are completely described by 4 MFs, i.e. the volume, surface area, integrated mean curvature, and Euler characteristic (or genus). Compared to the N-pt statistics, the MFs are more intuitional, they are easy to measure, and principally contain all orders of statistics simultaneously \citep{Mecke+94, Schmalzing+99}. Their application in studies of large-scale structure dates back to the 1990s \citep{Mecke+94}. Over the years, they have been applied with real surveys \citep{Kerscher+98, Hikage+03, Blake+14, WieEis17} to, e.g., test the Gaussianity of primordial fluctuations (with SDSS \citep{Hikage+03}) and construct standard rulers (with WiggleZ \citep{Blake+14}). Recently, they are newly proposed to probe theories of gravity \citep{PhysRevLett.118.181301}, mass of neutrinos \citep{PhysRevD.101.063515,liu2022probing} etc.

In this work, we study how peculiar velocities affect the morphological properties of large-scale structure as observed in redshift space. This is important both for the proposed applications of MFs and for studies of the peculiar velocitiy field itself. Previous work almost all focus on linear scales (studying the Kaiser's effect) without interpretating the morphological differences between redshift and real spaces \citep{1996ApJ...457...13M, Codis+13, 2020arXiv201208529L}. In this work, we focus on a comprehensive interpretation of the morphological differences between redshift and real spaces. As an attempt to extract the important information on non-linear structure formation, we go to the non-linear scales by utilizing large suite of N-body simulations. We also evaluate the cosmological constraints from the MFs and compare with the traditional 2-pt statistics of power spectrum.

\section{Calculations of the Minkowski Functionals}
\label{sec:sim}
 In this work, we use the Quijote simulations \cite{2020ApJS..250....2V} for our analysis of the MFs. The Quijote simulations are a large suite of N-body simulations generated for quantifying the information content of cosmological observables and training machine learning algorithms. The suite contains 44100 simulations spanning more than 7000 cosmological models in the $\left\{\Omega_{\mathrm{m}}, \Omega_{\mathrm{b}}, h, n_{s}, \sigma_{8}, M_{\nu}, w\right\}$ hyperplane. We use the subset of the Quijote simulations for the fiducial cosmology: $\Omega_m = 0.3175, \Omega_b = 0.049, h = 0.6711, n_s = 0.9624, \sigma_8 = 0.834, M_{\nu} = 0.0, w = -1$. The initial conditions are generated using 2nd-order Lagrangian perturbation theory at redshift z=127. Then they follow the gravitational evolution of $512^3$ dark matter particles in a cubic box with volume $1h^{-3}$Gpc$^3$ to $z = 0$ using the Gadget-III code \citep{10.1111/j.1365-2966.2005.09655.x}.

We use 300 simulations to estimate the theoretical means of the MFs in real and redshift spaces, from which we derive the morphological differences caused by the peculiar velocities, and estimate the errors or covariance matrix.
To forecast cosmological constraints from the MFs, we use the Fisher matrix technique  \citep{1997ApJ...480...22T, Jiang}. {We assume a Gaussian likelihood $\mathcal{L}$, with
\begin{equation}
-2\ln\mathcal{L}=[\bm{\mu}-\bar{\bm{\mu}}(\bm{\theta})]^{\rm T}\bm{C}^{-1}[\bm{\mu}-\bar{\bm{\mu}}(\bm{\theta})],
\end{equation}
where $\bm{\mu}$ represents the data vector of MFs (for some assumed observation), $\bar{\bm{\mu}}$ is the theoritical mean for $\bm{\mu}$ for a cosmological model with parameters specified by $\bm{\theta}$, which we calculate from simulations, and $\bm{C}$ is the covariance matrix. From the definition of the Fisher matrix,
\begin{equation}
F_{\alpha\beta}\equiv - \left< {\partial^2 \ln \mathcal{L}\over \partial \theta_\alpha\partial\theta_\beta} \right>,
\end{equation}
where the angular bracket denotes average over many realizations of the data, we obtain the following expression for $F_{\alpha\beta}$,
\begin{equation}
F_{\alpha\beta}=~{\partial\bm{\mu}^T\over\partial\theta_\alpha}\bm{C}^{-1}{\partial\bm{\mu}\over\partial\theta_\beta},
\end{equation}
where we have assumed a parameter-independent covariance matrix. For the estimation of the derivative,} we use 10 simulations for each parameter varied above or below its fiducial value.
 While for {$\bm{C}^{-1}$}, we multiply the inverse of the estimated covariance matrix by a factor of $(n-p-2)/(n-1)$ \citep{2007A&A...464..399H}, where $n$ is the number of samples and $p$ is the number of observables, to account for the bias in the inverse of the covariance matrix due to limited number of samples in its estimation. With this correction, we verified the numbers of simulations used to compute the derivatives and covariance matrix give convergent parameter constraints. Specifically, when we increase $n$ to 5000, the constraints differ by $\sim 10\%$. In the following, for cosmological constraints, we quote our most accurate estimation with $n=5000$.

When measuring the MFs, we adopt the common choice for the spatial pattern of large-scale structure as the excursion sets of the density field, i.e., regions with density above a given threshold. We construct the density field from the spatial distribution of dark matter particles using the cloud-in-cell mass-assignment scheme. To obtain a particle's position in redshift space $\vec{s}$, we adopt the distant-observer approximation such that $\vec{s}=\vec{r}+(1+z)\vec{v}_{\parallel}/H(z)$, where $\vec{r}$ is the position in real space, $z$ is the redshift, $\vec{v}_{\parallel}$ is the LOS component of the peculiar velocity, and $H(z)$ is the Hubble parameter. The dark matter density field is smoothed with a Gaussian window function with width $R_G$. We then measure the MFs for the smoothed field as a function of the density contrast $\delta$($\equiv\rho/\bar{\rho}-1$) used to specify the excursion sets. We find both the integral and differential methods of measuring the MFs numerically as developed in \citep{1997ApJ...482L...1S} give consistent results. In the following, we simply show our results obtained with the integral method.

\section{Results}
\label{sec:result}

\begin{figure*}[ht!]
\begin{center}
\includegraphics[width=1.0\textwidth,angle=0]{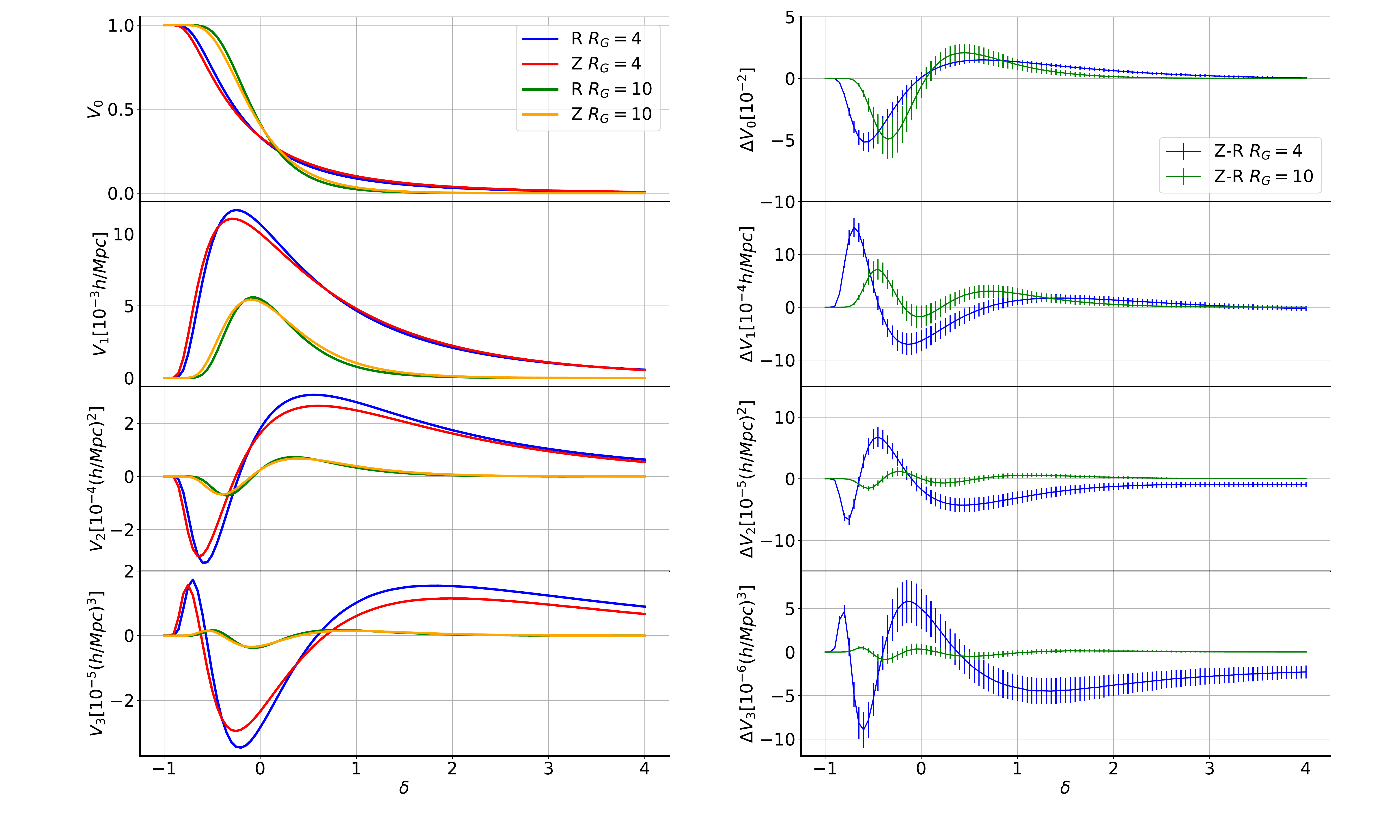}
\caption{ Left: The MFs of large-scale structure in real (labeled by ``R'') and redshift spaces (labeled by ``Z'') measured with $R_G=4h^{-1}\rm Mpc$ and at $z=0$ from the Quijote simulations for the fiducial model. $\delta$ is the density contrast used to define the excursion sets. The MFs for a larger smoothing scale $R_G=10h^{-1}\rm Mpc$ are also shown, for comparison. Right: Differences in the MFs between redshift and real spaces. Error bars are estimated using 300 simulations with volume $1h^{-3} \rm Gpc^3$, and have been enlarged 10 times here for ease of visualization.
\label{fig:1}}
\end{center}
\end{figure*}

 To find out the important morphological differences in large-scale sturcture caused by peculiar velocities, we measure and compare the MFs in real and redshift spaces at $z=0$ from the Quijote simulations for the fiducial model. 
We choose $R_G=4h^{-1}\rm Mpc$, which supresses shot noise while keeps most information of large-scale structure simultaneously. We also measure the MFs with $R_G=10h^{-1}\rm Mpc$ for a comparison of the (quasi-) linear and non-linear scales. 

We denote the four MFs as $V_i$ with $i=0,1,2,3$. In sequence, they represent the excursion sets' volume fraction, and surface area, integrated mean curvature, Euler characteristic per unit volume \citep[see e.g.,][for exact prefactors in the definitions]{1997ApJ...482L...1S}.
In Fig.~\ref{fig:1}, we show the measured MFs in real and redshift spaces in the left panel and their differences $\Delta V_i$ in the right panel. We display the results for the interval of $\delta\in[-1,4]$,
where the main features induced by peculiar velocities are captured while the signal-to-noise ratio for individual MF remains significant. Note error bars here are estimated using the real-space MFs measured from 300 simulations with volume $1h^{-3} \rm Gpc^3$. 
Since the error bars are so small, in the right panel of Fig.~\ref{fig:1}, we enlarge them by a factor of 10 for easier visualization.

At each smoothing scale, we find the curves of MFs in both real and redshift spaces share similar trends as a Gaussian random field, which have been well studied in the literature, see e.g., \cite{1997ApJ...482L...1S}.
However, deviations from the Gaussian case due to non-linear gravitational evolution are stronger in redshift space than in real space, consistent with the bigger r.m.s of the density in redshift space. 
While when comparing the two choices for $R_G$, we find both $V_i$ and $\Delta{V_i}$ have different amplitudes except for the normalized volume fraction with $i=0$, with smaller amplitudes for larger $R_G$. This is because smoothing erases structures with scales smaller than $R_G$. A larger $R_G$ erases more structures, thus smaller amplitudes for $V_i$ and $\Delta{V_i}$. We also notice the curves for the two choices of $R_G$ have similar trends except for $\Delta{V_2}$ and $\Delta{V_3}$ in the high density threshold regions. The difference is probablly due to the FOG effect which is important on non-linear scale (high density threshold regions), thus disappears with large $R_G$, but shows up only when $R_G$ is small enough. In the following, we focus on discussions of the $\Delta{V_i}$s with $R_G=4h^{-1}\rm Mpc$.

{$V_0$ is the volume fraction occupied by regions whose densities are above a density threshold specified by $\delta$. We find that the values of $V_0$ are larger when $\delta\gsim0$ and smaller when $\delta\lsim0$ in redshift space. That is, the volume fraction with density above an overdensity threshold becomes larger, while above an underdensity threshold becomes smaller. The latter is equivalent to that the volume fraction with density below an underdensity threshold becomes larger.}
This is consistent with a larger standard deviation of the density field in redshift space due to the Kaiser's effect, which is $\sim 2\%$ larger according to our measurement from simulations.
For underdense thresholds, our result indicates that the total volume of voids is larger in redshift space, that is, voids are more abundant and/or larger, consistent with findings by \cite{1996ApJ...470..160R} and \cite{2021MNRAS.500..911C}.

\begin{figure}[ht!]
\begin{center}
\includegraphics[width=1.0\linewidth,angle=0]{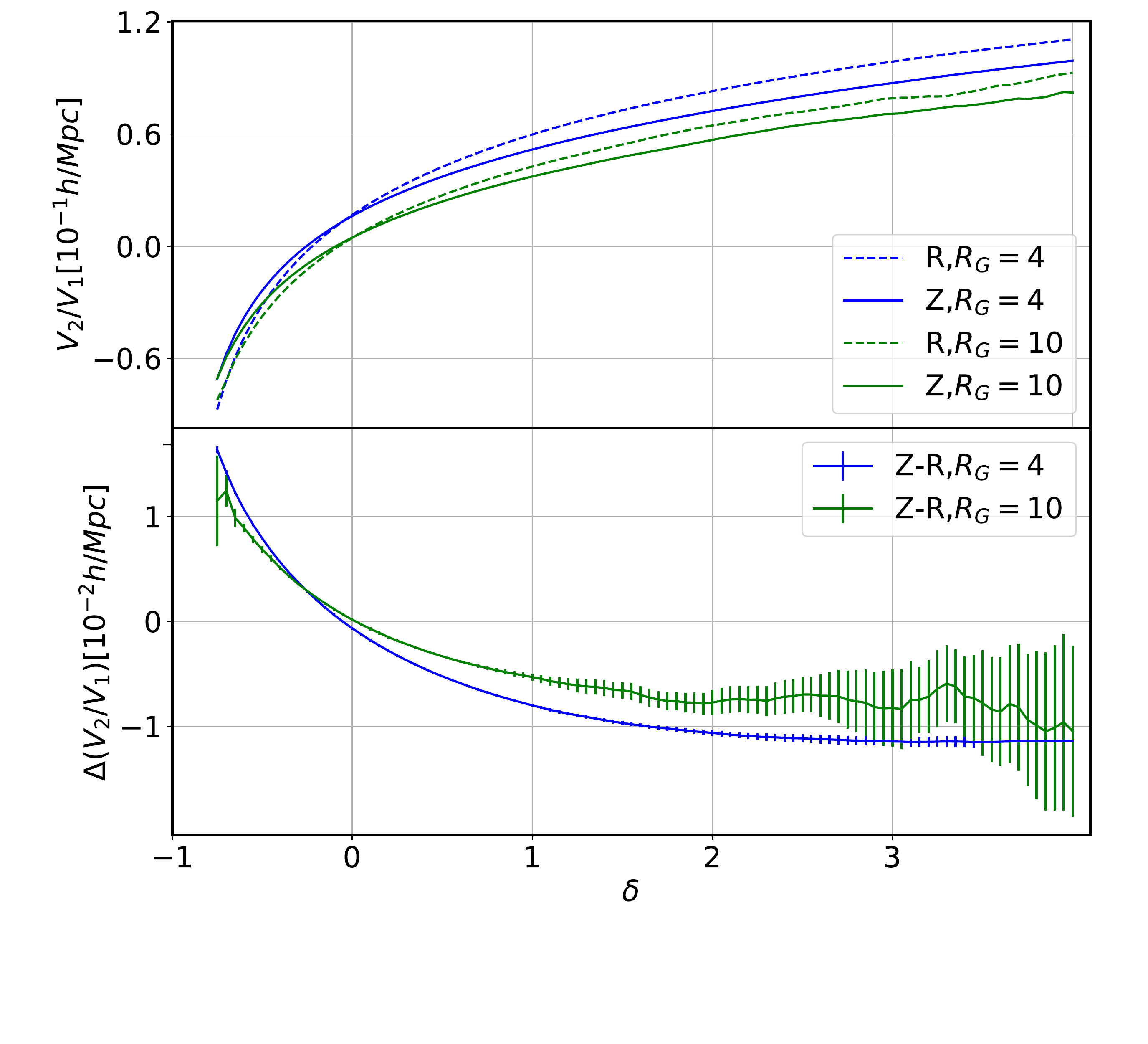}
\caption{Top: Ratio of $V_2$ to $V_1$ in redshift and real spaces as a function of $\delta$ for two smoothing scales $R_G=4h^{-1}\rm Mpc$ and $10h^{-1}\rm Mpc$. Bottom: The difference in this ratio between redshift and real spaces. 
\label{fig:2}}
\end{center}
\end{figure}

{$V_1$ is the surface area of the excursion sets or area of the isodensity contours. The values of $V_1$ are larger in redshift space except in the density range $-0.4\lsim\delta\lsim0.7$. If one assumes the excursion sets are all composed of isolated regions for a high enough density threshold, it is natural to expect that the change in their surface area follows that in the volume fraction they occupy. For a low enough density threshold, e.g. $\delta\lsim-0.4$, though one can still assume the isodensity contours are isolated, their enclosed regions are no longer the excursion sets but underdense regions with density below the threshold, whose volume fraction is therefore $1-V_0$. Thus $V_1$ becomes larger when $V_0$ becomes smaller. Besides, these isolated structures will no longer be statistically spherical in redshift space: overdense (underdense) regions for high (low) enough density threshold tend to be squashed (elongated) along the LOS. This departure from spherical shape also tends to cause an increase in the surface area. 
 However, the smaller $V_1$ in redshift space for $-0.4\lsim\delta\lsim0.7$ indicates deviations from the above perhaps simplest picture. One possibility is that with these thresholds, smaller structures merge, resulting in more large (fewer small) structures. Thus the surface area reduces while the volume increases. Another possibility is that holes on the isodensity contours become fewer in redshift space, leading to smaller surface area, consistent with changes in $V_3$ (see below).} 

 $V_2$ is the integration of the mean curvature over the surface area {of the isodensity contours. 
The sign of $V_2$ is determined by whether the surface is overall concave or convex.
 Since we define the positive direction of the surface pointing from lower to higher density region,} $V_2$ is negative for low density thresholds and positive for high density thresholds, with the transition taking place at $\delta\simeq-0.4$.
By comparing the curves of $V_2$ in real and redshift spaces, we find that the absolute value of $V_2$ is smaller in redshift space when $-0.6\lsim\delta\lsim-0.4$ and $\delta\gsim-0.2$, while larger elsewhere. 
From its definition, $V_2$ can be affected by both changes in the mean curvature and those in the surface area. To separate the effect from the latter, {which we have already obtained,} we introduce the ratio of $V_2$ to $V_1$, {and focus on discussions of its changes caused by the redshift space distortions}. This is the surface-area-weighted average of the mean curvature (hereafter, ``average curvature'' for short) for the isodensity contours. 
In Fig.~\ref{fig:2}, we plot $V_2/V_1$ and the difference in it between redshift and real spaces.

Before we discuss the difference in $V_2/V_1$ caused by the peculiar velocities, let us first look at this ratio for an ellipsoid with semi-axes $(a,a,\lambda a)$. When $\lambda$ is fixed, both $V_1$ and $V_2$ increase with $a$, but $V_2/V_1$ decreases. While when $a$ is fixed, $V_2/V_1$ increases monotonically with $|\lambda-1|$, and reaches its minimum at $\lambda=1$ \citep[see][for more details]{sors2004integral}. {As a summary, enlarging structures's sizes overall leads to smaller $|V_2/V_1|$, while changing their shapes to be non-spherical leads to larger $|V_2/V_1|$.}

For $R_G=4h^{-1}\rm Mpc$, we find that the average curvature $V_2/V_1$ is smaller in redshift space {when $\delta\gsim0$, while larger when $\delta\lsim0$.}
Considering the sign of $V_2/V_1$, we can deduce $|V_2/V_1|$ is smaller {except in a narrow threshold range $-0.2\lsim\delta\lsim0$}. Smaller value of $|V_2/V_1|$ indicates bigger underdense regions (voids) and overdense regions (halos) in redshift space, consistent with our findings from the changes in $V_0$. {While for $-0.2\lsim\delta\lsim0$, positive average curvature gets bigger in redshift space.}
Isotropic overdense structure in real space is squashed along the LOS forming an ellipsoid with $\lambda<1$, which leads to a larger average curvature in redshift space, just as what we find {for $-0.2\lsim\delta\lsim0$} in Fig~\ref{fig:2}.

 

The Euler characteristic $V_3$ measures the connectedness of the excursion sets, which equals the number of isolated structures minus the number of holes per unit volume \citep[see e.g.][]{1996ApJ...463..409M, 1999IAUS..183..210S}. With $R_G=4h^{-1}\rm Mpc$, the negative $V_3$ for $-0.6\lsim\delta\lsim0.6$ indicates the excursion sets are more connected, that is, there are fewer disjoint regions but more holes. The positive $V_3$ for higher or lower density thresholds indicates the opposite case. We find in redshift space, the excursion sets are more connected with $\Delta{V_3}<0$ for $-0.8\lsim\delta\lsim-0.4$ and $\delta\gsim0.4$, while less connected with $\Delta{V_3}>0$ elsewhere. 
With a larger smoothing scale $R_G=10h^{-1}\rm Mpc$ or above, we find $\Delta{V_3}>0$ for high density thresholds, which is different from the small smoothing scale results.


\begin{figure}[ht!]
\begin{center}
\includegraphics[width=0.95\linewidth,angle=0]{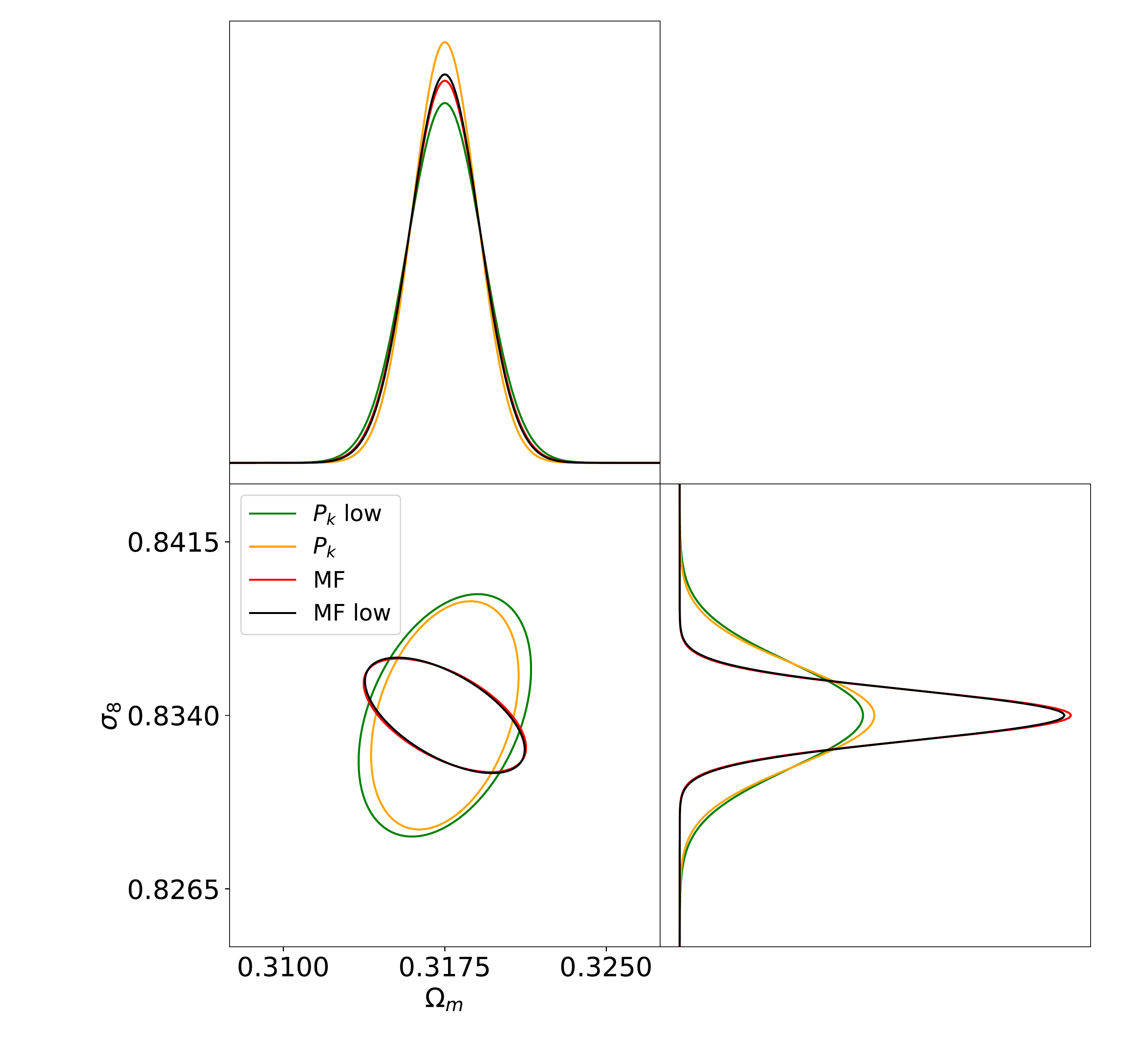}
\caption{ {$68\%$ confidence regions and marginalized likelihood functions for $\Omega_{\rm m}$ and $\sigma_8$ obtained from the redshift-space MFs and power spectrum (specifically the combination of monopole, quadrupole and hexadecapole).
{ The red (MFs) and orange (power spectrum) contours represent the constraints forecasted using the fiducial Quijote simulations. The black (MFs) and green (power spectrum) contours are results using the low-resolution simulations from the Quijote suite.} }
\label{fig:3}}
\end{center}
\end{figure}
 
From Fig.~\ref{fig:1}, we can see that the fractional changes in the MFs induced by peculiar velocities are on the order of $\sim10\%$ when $R_G=4h^{-1}\rm Mpc$, and judging from the error bars, they can be detected with significant $S/N$. {Thus, the MFs are very sensitive to redshift space distortion. They may provide a promising way to study the peculiar velocity field, which deserves further explorations in future work.
Here, we perform a more straightforward study of the cosmological constraints obtainable from the MFs in redshift space, by using the Fisher matrix technique and the Quijote simulation data.}
We choose the matter density parameter $\Omega_m$ and $\sigma_8$, the amplitude of the linear density fluctuations on a scale of $8h^{-1}\rm Mpc$, as representatives, which most directly influence the redshift-space clustering. Our results are shown in Fig.~\ref{fig:3} (red contours), where we have chosen $R_G=4h^{-1}\rm Mpc$ and {$21$ density threshold bins for each order of the MFs}. For comparison, the constraints from the redshift-space matter power spectrum, specifically the combination of its monopole, quadrupole and hexadecapole are also shown (orange contours), {where we have chosen $21\ k$ bins up to $k_{\rm max}=0.5\ h$Mpc$^{-1}$.} As can be seen, in redshift space, the MFs give relative stronger constriants than power spectrum, probablly due to the non-linear statistical information contained in the MFs. 

{We note these constraints} are estimated for the simulated dark matter distribution with number density $0.13\ h^3$Mpc$^{-3}$. This tracer number density clearly sounds optimistic for current galaxy surveys. {For a more realistic estimation, we use the low-resolution simulations in the Quijote suite, which have a dark matter density of $0.017\ h^{3}\rm Mpc^{-3}$, roughly the expected number density for the BGS galaxies that the currently ongoing DESI survey is about to observe \citep{2016arXiv161100036D}. The results are shown as the black (MFs) and green (power spectrum) contours in Fig~\ref{fig:3}. One can see that this reduction of tracer number density has little impact on the constraints from the MFs with $R_G=4h^{-1}\rm Mpc$, while it does make the constraints from the power spectrum a bit worse, indicating shot noise is sub-dominant on the involved scales. For $\sigma_8$, we find the constraint from the MFs is now better by $\sim 50\%$ than that from the power spectrum.}

\section{Conclusions}
\label{sec:con}

Using large suite of N-body simulations, we have studied the peculiar velocity induced features in the four MFs of large-scale structure. With a focus on non-linear scale, we present detailed interpretation for the morphological changes of LSS in redshift space. 
With the Quijote simulations, we perform a Fisher matrix analysis for the cosmological constraints from the MFs, and compare with those from the power spectrum in redshift space, and find the MFs can give overall better constraints.  {Specifically, for galaxy number density similar to the DESI BGS galaxies, we find the constraint from the MFs on $\sigma_8$ is better by $\sim 50\%$ than from the power spectrum. We find this number density is sufficient for the constraints from the MFs to be converged regarding shot noise.}
{Note these calculations are performed for simulated cubic boxes with periodic boundary conditions. When applied to real surveys, the irregular survey mask and non-periodic boundary should be taken into account (see e.g., \citep{Blake+14,appleby2022minkowski} for how to measure the MFs with survey masks).}
{We also notice other systematic effects, e.g. the bias, pixelization effects, might change the morphological features induced by peculiar velocities, and leave a quantitive analysis for future work.}  This work highlights and paves the way for the applications of MFs of large-scale structure in real galaxy surveys, which we are currently pursuing.

\begin{acknowledgments}
We thank the Quijote team for sharing their simulation data, and specially thank Francisco Villaescusa-Navarro for instructions on how to use the data. We thank {an anonymous referee}, Yu Yu, Pengjie Zhang, Gong-Bo Zhao for valuable suggestions and comments, and thank Ziwei Jia for help with some of the mathematical calculations. 
This work is supported by the National Natural Science Foundation of China Grants No. 12173036, 11773024, 11653002, 11421303, by the National Key R\&D Program of China Grant No. 2021YFC2203100, by the China Manned Space Project Grant No. CMS-CSST-2021-B01, by the Fundamental Research Funds for Central Universities Grants No. WK3440000004 and WK3440000005, and by the CAS Interdisciplinary Innovation Team.

\end{acknowledgments}

\bibliographystyle{apsrev4-2}
\bibliography{reference}

\end{document}